\documentclass[sigplan,screen]{acmart}

\usepackage{amsmath}
\usepackage{multirow}

\AtBeginDocument{%
  \providecommand\BibTeX{{%
    \normalfont B\kern-0.5em{\scshape i\kern-0.25em b}\kern-0.8em\TeX}}}

\begin{document}
\title{MUSE2020 Challenge Report}
\author{Ruichen Li}
\email{ruichen@ruc.edu.cn}
\affiliation{%
  \institution{School of Information Renmin University of China}
  \city{Beijing}
  \country{China}
}

\author{Jingwen Hu}
\email{2015201993@ruc.edu.cn}
\affiliation{%
  \institution{School of Information Renmin University of China}
  \city{Beijing}
  \country{China}
}

\author{Shuai Guo}
\email{2016202112@ruc.edu.cn}
\affiliation{%
  \institution{School of Information Renmin University of China}
  \city{Beijing}
  \country{China}
}

\author{Jinming Zhao}
\email{zhaojinming@ruc.edu.cn}
\affiliation{%
  \institution{School of Information Renmin University of China}
  \city{Beijing}
  \country{China}
}

\begin{abstract}
  This paper is a brief report for MUSE2020 challenge. We present our solution for Muse-Wild sub challenge. The aim of this challenge is to investigate sentiment analysis method in real-world situation. Our solutions achieve the best CCC performance of 0.4670, 0.3571 for arousal, and valence respectively on the challenge validation set, which outperforms the baseline system with corresponding CCC of 0.3078 and 1506.
\end{abstract}

\maketitle

\section{Introduction}
Multimodal Sentiment Analysis is a significant task which helps people leverage daily data better from visual, textual and acoustic modalities. With the help of this task, amounts of systems are built incorporating conversational agents\cite{Cowie2002Emotion}, education tutoring\cite{Conati2002Probabilistic} and etc.

Muse-Wild, a sub-challenge of MuSe(Multimodal Sentiment Analysis in Real-life Media), argues participants to predict the level of affective dimensions including arousal(a affective activation) and valence(a measure of pleasure), which are described by dimensional theory, one of the most important affective computing theories. Some significant emotion recognizing systems are constructed based on the theory.

Previous studies on multimodal sentiment analysis have applied LSTM+Self-Att and EndYou model, which performs well in this task. In addition, (introduce main work in this competition).

Our contributions to the challenge in this paper are from two aspects:
\begin{itemize}
    \item We investigate several efficient features differing from given features of MuSe from acoustic, visual and textual modalities.
    \item Our ensemble model achieve the best CCC performance of 0.4121 for arousal and valence, which outperforms the baseline systems with corresponding CCC of 0.2047 for arousal and valence.
\end{itemize}

\section{Multimodal Features}
\subsection{Textual Features}
We tried Bert\cite{devlin2018bert}, Albert\cite{Lan2019ALBERT}, Glove models to encode the text respectly. Specifically, pretrained models are used to extract text features, then we averaged the word or character features the timestamps of which overlapped a certain 250ms frame as a method of aligning. We refer there features as “bert\_cover”, “albert\_cover” and “glove\_cover”.
\subsection{Audio Features}
We tried egemaps \cite{egemaps} LLD(Low-level Descriptors) set in describing the acoustic emotion features. In addition, self-supervised model like wav2vec\cite{wav2vec}, is also used for improving acoustic model training. In order to get better results, the wav2vec model is pretrained on Librispeech dataset in advance. We refer these features the same as its name.

\subsection{Visual Features}
We use DenseFace model and VggFace feature as visual input. The DenseFace model is pretrained on FER+ dataset \cite{FER} use same structure and finetuning strategy proposed in \cite{AVEC2019}.We extract features from last mean pooling layer of the finetuned model and refer the features as "denseface". we use vgg16 model which is pretrained on Vggface dataset \cite{vggface} and finetuned the same way as denseface, vggface feature is refered as "vggface".

\section{Proposed Model}
In this section we will describe our proposed method in detail. We use combination of different features described in section 2 to solve Muse-Wild challenge.

Here we formulate the question, each video is framed into segments of 250ms, we have

\begin{equation}
    \begin{aligned}
        v_{i} &= [X, Y], \\
        X &= \{x_{i}^{j}\},  \\
        Y &= \{y_{j}\}, \quad i=1,2,...K, j=1,2,...t
    \end{aligned}
\end{equation}

K is the number of different types of feature for model input, and t is the number of segments in a video. Each $y_{j}$ is a number ranged in [-1, 1], which could be arousal or valence, $x_{i}^{j}$ means a feature vector for $i^{th}$ feature and $j^{th}$ timestamp.

For each different input feature, we firstly concatenate them together by timestamps,
\begin{equation}
z_{j} = concat([x_{1}^{j}, x_{2}^{j}, ..., x_{K}^{j}]
\end{equation}
then a fully-connected layer with RELU activation maps the features into a embedding space, we use a LSTM module to encode time sequential information of the input sequence. Finally 2 fully-connected layer is used for label regression.
\begin{equation}
    \hat{z_{j}} = RELU(W*z_{j}+b)
\end{equation}

\begin{equation}
    \begin{aligned}
        h &= LSTM([\hat{z_{1}}, \hat{z_{2}}, ..., \hat{z_{t}}] \\
        h &= [h_{1}, h_{2}, ..., h_{t}]
    \end{aligned}
\end{equation}

\begin{equation}
    \hat{y_{j}} = fully\_connect(h_{j}) 
\end{equation}

We calculate MSE(mean square error) between each $y_{j}$ and $\hat{y_{j}}$ as supervise signal during training.
\begin{equation}
    loss = \sum_{j=1}^{t}{(y_{j} - \hat{y_{j}})^2}
\end{equation}

\section{Experiments}
\subsection{Expreiment Setup}
As described in section3, we use LSTM as our encoder, the number of layers is set to be 1 and the number of hidden units is optimized for different input features. Adam optimizer is applied to optimize the model, and we set the max time step to be 100 and the dropout rate to be 0.5.

\subsection{Expreiment Result}
As shown in the tabel 1, in uni-modal experiments, lld feature has the best validation performance on arousal, and bert feature works best on valence. In multi-modal experiments, the feature combination of lld, wav2vec, denseface, au and bert feature has the best performance, which reached CCC of 0.4670 on arousal and 0.3571 on valence.

\begin{center}
\begin{table}[]
\begin{tabular}{|c|l|c|c|}
\hline
\multicolumn{2}{|c|}{Features}                               & arousal         & valence         \\ \hline
\multirow{5}{*}{Uni-modal}   & lld                           & \textbf{0.3841} &      0.03           \\ \cline{2-4} 
                             & wav2vec                       & 0.3092          & 0.1737          \\ \cline{2-4} 
                             & denseface                     & 0.2954          & 0.0524          \\ \cline{2-4} 
                             & vggface                       & 0.2996          & 0.0852          \\ \cline{2-4} 
                             & bert\_base           & 0.1643          & \textbf{0.3131} \\ \hline
\multirow{4}{*}{Multi-modal} & lld-bert-glove                & 0.3378          & 0.3447          \\ \cline{2-4} 
                             & wav2vec-bert\_base-glove      & 0.3440          & 0.3556          \\ \cline{2-4} 
                             & lld-wav2vec-denseface-au-bert & \textbf{0.4670} & \textbf{0.3571} \\ \cline{2-4} 
                             & lld-wav2vec-vggface-au-bert     & 0.4514          & 0.3153          \\ \hline
\end{tabular}
\caption{Experiment Result on Validation Set}
\end{table}
\end{center}

\section{Conclusion}
In this paper, we explore different efficient deep learning features from acoustic, visual and textual modalities in real-world sentiment analysis task, our proposed model reached CCC of 0.4670 on arousal and 0.3571 on valence on validation set.

\bibliographystyle{ACM-Reference-Format}
\bibliography{sample}

\end{document}